\documentclass[12pt,amsmath]{iopart}

\usepackage{graphicx}

\begin{document}

\title{Dichotomy in the $T$-linear resistivity in hole-doped cuprates}

\author{N. E. Hussey$^1$, R. A. Cooper$^1$, Xiaofeng Xu$^{1,2}$, Y. Wang$^{1,3}$, B. Vignolle$^4$ and C. Proust$^4$}
\address{$^1$ H. H. Wills Physics Laboratory, University of Bristol, Tyndall Avenue, Bristol, BS8 1TL, UK}
\address{$^2$ Department of Physics, Hangzhou Normal University, Hangzhou 310036, China}
\address{$^3$ State Key Laboratory for Mesoscopic Physics and School of Physics, Peking University, Beijing 100871, China}
\address{$^4$ Laboratoire National des Champs Magn\'{e}tiques Intenses (CNRS), Toulouse 31400, France}

\begin{abstract}
From analysis of the in-plane resistivity $\rho_{ab}(T)$ of La$_{2-x}$Sr$_x$CuO$_4$, we show that normal state transport in overdoped cuprates can be delineated into two regimes in which the electrical resistivity varies approximately linearly with temperature. In the low temperature limit, the $T$-linear resistivity extends over a very wide doping range, in marked contrast to expectations from conventional quantum critical scenarios. The coefficient of this $T$-linear resistivity scales with the superconducting transition temperature $T_c$, implying that the interaction causing this anomalous scattering is also associated with the superconducting pairing mechanism. At high temperatures, the coefficient of the $T$-linear resistivity is essentially doping independent beyond a critical doping $p_{\rm crit}$ = 0.19 at which the ratio of the two coefficients is maximal. Taking our cue from earlier thermodynamic and photoemission measurements, we conclude that the opening of the normal state pseudogap at $p_{\rm crit}$ is driven by the loss of coherence of anti-nodal quasiparticles at low temperatures.
\end{abstract}




\submitto{Phil. Trans. R. Soc. A}

\maketitle

\section{Introduction}

Although the empirical phase diagram of high-$T_c$ cuprates has been drawn countless times, the precise form (i.e. temperature vs. doping curve) of the various temperature scales is still a relatively subjective exercise. Only the N\'{e}el temperature $T_{\rm N}$ and $T_c$, defining the onset of long-range antiferromagnetic (AFM) order and long-range superconducting phase coherence respectively, are well established experimentally. Other scales are either drawn arbitrarily or follow entirely different trajectories depending on the type of experimental probe employed. The \lq coherence' temperature $T_{\rm coh}$, often used to describe the emergence of a conventional Fermi-liquid ground state at high doping, is one such example, though by far the most controversial temperature scale is the pseudogap onset temperature $T^\ast$.

This controversy was articulated neatly in a recent review by Norman, Pines and Kallin (Norman {\it et al.} 2005) in which they summarized the three most common scenarios for $T^\ast$($p$), $p$ being the doped hole concentration. Spectroscopic probes like angle-resolved photoemission (ARPES), scanning tunneling microscopy (STM) and Raman scattering find that $T^\ast$ gradually merges with $T_c$ beyond optimal doping (H\"{u}fner {\it et al.} 2008). In this scenario, the pseudogap state is often regarded as a precursor to superconductivity involving strong pairing correlations without long-range phase coherence. Thermodynamic probes such as specific heat and magnetic susceptibility on the other hand find a $T^\ast$($p$) line that cuts the top of the superconducting dome and vanishes at some critical doping level around $p_{\rm crit}$ = 0.19 (Tallon \& Loram 2001). This delineation of the two temperature scales is supported by polarized neutron scattering (Fauqu\'{e} {\it et al.} 2006) and polar Kerr effect studies (Xia {\it et al.} 2008) and suggests that the pseudogap state is some kind of ordered state that competes with superconductivity by removing spectral weight which is never fully recovered upon entering the superconducting phase: $p_{\rm crit}$ is often then interpreted as a critical end-point with (quantum) critical fluctuations associated with this ordered state influencing the physical properties over a large funnel-shaped region in the ($T, p$) phase diagram in which marked deviations from conventional Landau Fermi-liquid behaviour are manifest.

As most transport coefficients (e.g. resistivity, Hall resistance, thermopower etc...) vanish at the onset of bulk superconductivity, the outcome of transport probes has tended to align with the third and final scenario in Norman's review in which the $T^\ast$($p$) line simply terminates at the top of the superconducting dome. As such, transport specialists have tended to \lq sit on the fence' on this issue, though the observation of a broad funnel-shaped region of $T$-linear resistivity has been widely interpreted, in line with similar observations in heavy fermion systems such as YbRh$_2$Si$_2$ (Custers {\it et al.} 2004) in terms of quantum criticality. The location of this putative quantum critical point (QCP) however, as well as unamgibuous signatures of quantum criticality in the transport behaviour, has remained elusive, largely due to the high upper critical field $H_{c2}$ values in hole-doped cuprates that restrict access to the important limiting low-temperature region below $T_c(p)$.

In a recent report (Cooper {\it et al.} 2009), we employed a combination of persistent and pulsed high magnetic fields to expose the normal state of La$_{2-x}$Sr$_x$CuO$_4$ (LSCO) over a wide doping and temperature range and studied the evolution of $\rho_{ab}(T)$ with carrier density, from the slightly underdoped ($p$ = 0.15) to the heavily overdoped ($p$ = 0.33) region of the phase diagram. Rather than collapsing to a singular (critical) point, the region of $T$-linear resistivity in LSCO was found to fan out at low temperatures and dominate the low-$T$ response over a very wide doping range. This anomalous or \lq extended' critical region was wholly unexpected and contrasts markedly with what is observed in YbRh$_2$Si$_2$ (Custers {\it et al.} 2004) and in other candidate quantum critical systems (Paglione {\it et al.} 2003).  Our analysis also revealed that the magnitude of the $T$-linear term scaled monotonically with $T_c$ on the strongly overdoped side but saturated, or was maximal, at a critical doping level $p_{\rm crit} \sim$ 0.19 at which superconductivity itself is most robust. The observation of a singular doping concentration in LSCO close to $p$ = 0.19 at which a bulk transport property undergoes a fundamental change at low $T$ lends support to the claim that the pseudogap temperature $T^{\ast}$ or energy scale $\Delta_g$ vanishes inside the superconducting dome, rather than at its apex.

In this article, we explore this issue further by examining the temperature derivative of the normal state electrical resistivity of LSCO across the same doping range. By focussing on the temperature range between the zero-field $T_c$ and 300 K, we show that the $T$-linear resistivities at high and low temperatures have different gradients, different doping dependencies and by inference, different origins. In the low temperature limit, the $T$-linear resistivity appears to result from scattering processes that are associated with the superconducting pairing mechanism. The high-temperature $T$-linear resistivity, by contrast, is attributed to the onset of incoherence, initially of quasiparticles located near the anti-nodes. At $p$ = $p_{\rm crit}$, this coherence temperature falls below $T_c$, linking the loss of coherence to the opening of the pseudogap itself. Finally, kinks in the temperature derivative of $\rho_{ab}(T)$ near $T_c$ define an additional temperature scale $T_F$ in the phase diagram of overdoped cuprates associated with the onset of anomalous superconducting phase fluctuations previously seen in Nernst measurements at lower doping (Xu {\it et al.} 2000). Separation of these two temperature scales ($T^{\ast}$ and $T_F$) in the electrical resistivity allows us finally to unify the different scenarios for the phase diagram of hole-doped high-$T_c$ cuprates.

\section{Experiment and Results}

Single crystals of LSCO with various Sr concentrations were grown using the travelling-solvent floating-zone method and annealed in varying partial pressures of oxygen in order to optimize homogeneity of the oxygen concentration and thus avoid phase separation into hole-rich and hole-poor regions. Crystallographic axes were identified using a Laue camera, and electrical contacts mounted onto individual crystals in such a way as to avoid voltage contamination along the $c$-axis. The low-field ($\mu_0 H <$ 18 Tesla) magnetoresistance measurements were carried out using a standard superconducting magnet whilst the high-field measurements up to 60 Tesla were performed in a standard $^4$He cryostat at the LNCMI-T pulsed-field facility in Toulouse.

\begin{figure}
\begin{center}
\includegraphics [width=15cm, bb=0 0 1046 742]{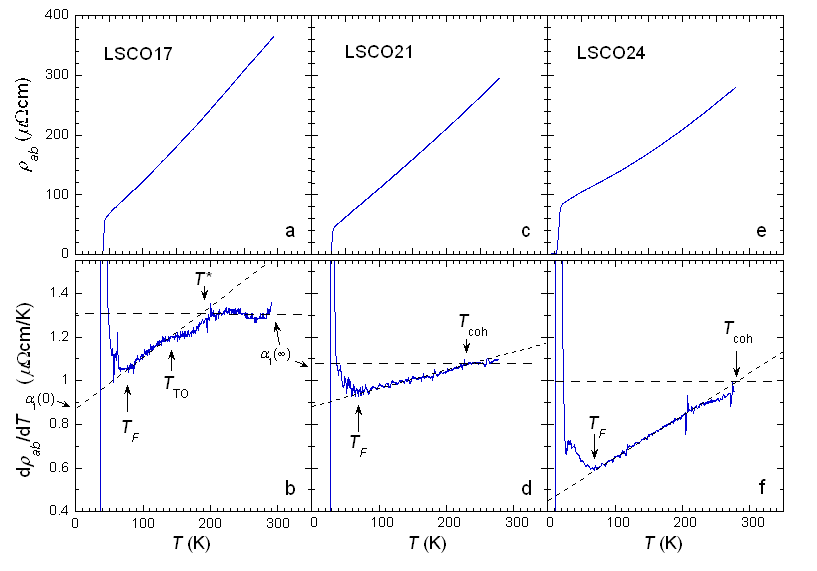}
\caption {Top panels: Temperature dependent zero-field resistivity curves for La$_{2-x}$Sr$_x$CuO$_4$ with $x$ = (a) 0.17, (c) 0.21 and (e) 0.24. Bottom panels: Corresponding temperature derivative curves for $x$ = (b) 0.17, (d) 0.21 and (f) 0.24. The onset temperature for superconducting fluctuations is labelled $T_F$, the tetragonal to orthorhombic structural transition by $T_{\rm TO}$, the opening of the pseudogap by $T^{\ast}$ and the onset of quasiparticle incoherence by $T_{\rm coh}$. See text for further details.}\label{Figure1}
\end{center}
\end{figure}

Figure 1 shows resistivity data for three representative dopings, $x$ = 0.17, 0.21 and 0.24 (denoted LSCO17 LSCO21 and LSCO24 respectively) with the corresponding temperature derivatives plotted in the lower three panels. Whilst the resistivity curves appear smooth and featureless, the derivatives themselves reveal a surprising degree of structure that can be assigned to distinct physical phenomena each with a characteristic temperature scale. The marked increase in the temperature derivative at $T$ = $T_F$ can be clearly identified as the onset of superconducting fluctuations that cause a downturn in the normal state resistivity. As we shall show later, at doping levels where such measurements have been performed, this temperature coincides with the onset of a positive Nernst coefficient that was shown by Ong and co-workers to be generated by short-lived vortex excitations above the bulk superconducting transition (Xu {\it et al.} 2000).

\begin{figure}
\begin{center}
\includegraphics [width=10.0cm, bb=0 0 1046 1360]{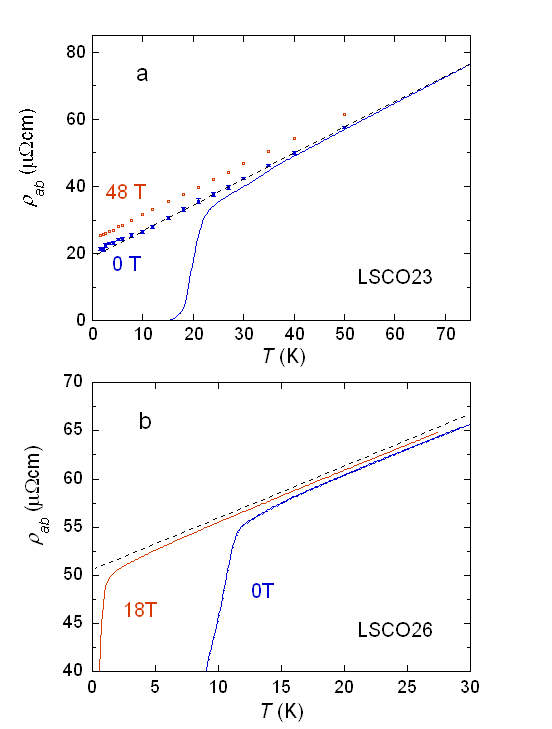}
\caption {(a) Zero-field $\rho_{ab}(T)$ of one LSCO23 crystal (solid blue line), plotted together with the extrapolated values of $\rho$(0) (blue closed squares - see text) and of  $\rho$($\mu_0 H$ = 48 T) (brown open squares) for the various temperatures indicated. (b) $\rho_{ab}(T)$ of one LSCO26 single crystal in zero-field (blue line) and a magnetic field of 18 Tesla applied parallel to the $c$-axis (brown line). The dashed black lines in both panels serve to highlight the preservation of the linear $T$-dependence of the resistivity to lower temperatures as superconductivity is destroyed.}
\label{Figure2}
\end{center}
\end{figure}

Above $T_F$, d$\rho_{ab}$/d$T$ rises linearly with increasing temperature (bar the kink at $T_{\rm TO}$ in lower doped samples to be discussed below) until finally it plateaus at a value we label $\alpha_1(\infty)$, the high-temperature limit of the $T$-linear resistivity. (For the highest doping levels, this limit is reached asymptotically). Extrapolation of d$\rho_{ab}$/d$T$ to $T$ = 0 results in a second value of the $T$-linear resistivity that we label $\alpha_1(0)$. The reliability of this extrapolation and the recovery of $T$-linear resistivity as $T \rightarrow 0$ is highlighted in Figure 2 for two different dopings (a) LSCO23 and (b) LSCO26. In LSCO23, a (pulsed) field of 48 Tesla is sufficient to recover the normal state (Cooper {\it et al.} 2009). The temperature dependence of $\rho(\mu_0 H$ = 48 T) is plotted as open squares in Fig. 2 and compared with estimates (closed squares) of $\rho(0)$, the resistivity extrapolated to zero-field assuming a simple quadratic field dependence of the magnetoresistance. Both exhibit a clear dominant $T$-linear dependence below 60 K. In LSCO26, a field of 18 Tesla is found to extend the range of the $T$-linear resistivity down to at least 5 K, below which residual superconductivity drives the resistivity towards zero. Note that the preservation of the $T$-linear resistivity above $H_{c2}$ is counter to recent claims of field-induced quantum criticality via the observation of a $T^2$ $c$-axis resistivity in overdoped Tl$_2$Ba$_2$CuO$_{6+\delta}$ (Tl2201) above $H_{c2}$ (Shibauchi {\it et al.} 2008). In that particular configuration, $c$-axis magnetoresistance is large and strongly $T$-dependent (French \& Hussey 2008) and as a consequence, limits one's ability to extract the intrinsic ($ab$-plane) response.

In our previous paper (Cooper {\it et al.} 2009), we showed that for $p > 0.17$, all $\rho_{ab}(T)$ curves for 0 $< T \leq$ 200 K could be fitted to the expression

\begin{equation}
\rho_{ab}(T) = \alpha_0 + \alpha_1(0) T + \alpha_2 T^2
\end{equation}

\noindent
where $\alpha_0$ is the residual resistivity, $\alpha_1(0)$ the coefficient of the low-$T$ $T$-linear term and $\alpha_2$ the coefficient of the quadratic term.

This form of resistivity is self-evident from the temperature derivatives displayed in Fig. 1, in particular the linear slope of d$\rho_{ab}$/d$T$ above $T_F$. A similar expression is also found to describe $\rho_{ab}(T)$ in overdoped Tl2201 at low temperatures (Mackenzie {\it et al.} 1996, Proust {\it et al.} 2002). Note that the observed form of d$\rho_{ab}$/d$T(T)$ is inconsistent with the single component analysis, i.e. $\rho_{ab}(T) = \alpha_0 + \alpha_n T^n$, advocated by some groups (Manako {\it et al.} 1992, Naqib {\it et al.} 2003) as according to the latter, d$\rho_{ab}$/d$T(T)$ should exhibit strong curvature and go to zero at $T$ =0. For the highest $x$-values ($p \geq 0.24$), d$\rho_{ab}$/d$T(T)$ does exhibit downward curvature, but only at high temperatures. The latter can be interpreted either as a tendency towards resistivity saturation (Hussey 2003, 2008, Cooper {\it et al.} 2009) or as the development of a non-integer power law in $\rho_{ab}$($T$) at high temperatures, due e.g. to development of ferromagnetic critical fluctuations near the apex of the superconducting dome (Kopp {\it et al.} 2007, Sonier {\it et al.} 2009).

The form of $\rho_{ab}(T)$ described in Eqn. (1) is also consistent with the form of the in-plane transport scattering rate $\Gamma$ extracted from angle-dependent magnetoresistance (ADMR) measurements in overdoped Tl2201 (Abdel-Jawad {\it et al.} 2006, French {\it et al.} 2009). In particular, $\Gamma$ is found to be composed of two components with different $T$ and momentum ({\bf k}) dependencies:  one ($\gamma_{\rm iso}$) isotropic and quadratic in $T$, the other ($\gamma_{\rm aniso}$) anisotropic, maximal near the saddle points at ($\pi$, 0) and proportional to temperature. The fact that the two $T$-dependent components in $\Gamma$($T$, {\bf k}) and $\rho_{ab}(T)$ are {\it additive} implies the presence of two distinct, independent quasiparticle scattering processes that coexist {\it everywhere} on the cuprate Fermi surface. This contrasts with models, e.g. based on hot spots (Stojkovic \& Pines 1997) or cold spots (Ioffe \& Millis 1998), in which different regions of {\bf k}-space have different relaxation rates, since in these cases, the lifetimes ought to be additive.

\section{Discussion}

\begin{figure}
\begin{center}
\includegraphics [width=11.0cm, bb=0 0 592 720]{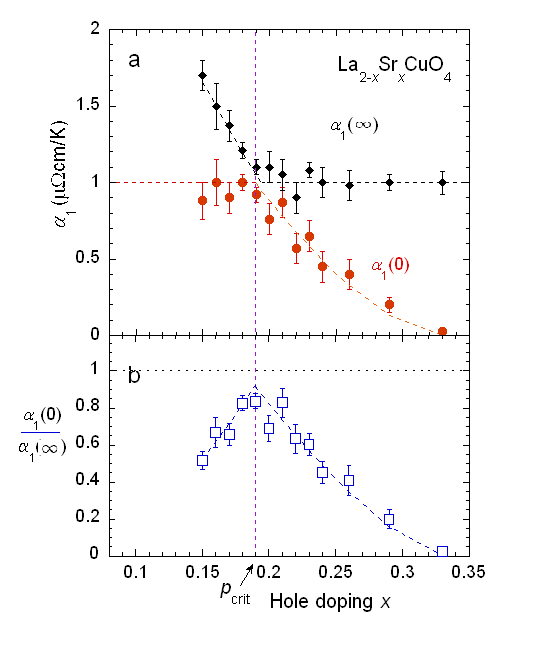}
\caption {(a) Doping dependence of $\alpha_1(\infty)$ (black diamonds) and $\alpha_1(0)$ (red circles) for La$_{2-x}$Sr$_x$CuO$_4$. The thin dashed lines are guides to the eye; the thick dashed line signifies $p$ = $p_{\rm crit}$. (b) Ratio of $\alpha_1(0)$/$\alpha_1(\infty)$ as a function of Sr concentration. Again, the dashed lines are guides to the eye.}\label{Figure3}
\end{center}
\end{figure}

Figure 3 summarizes the doping dependence of $\alpha_1(\infty)$ (black diamonds) and $\alpha_1(0)$ (red circles), collated from a large number of LSCO crystals with closely spaced Sr concentrations (including analysis of data for 0.15 $< x <$ 0.21 reported by Ando {\it et al.} 2004). For $x >$ 0.30,   $\alpha_1(0) \sim$ 0 and the low-$T$ resistivity is strictly quadratic (Nakamae {\it et al.} 2003, 2009). With decreasing $x$, $\alpha_1(0)$ grows rapidly, attaining a maximum value of around 1 $\mu\Omega$cm/K at $p_{\rm crit}$ = 0.19 $\pm$ 0.01. It is interesting to note that this anomalous, distinctly non-Fermi-liquid dependence of the electrical resistivity occurs in a regime of the phase diagram where coherent fermionic quasiparticles, as revealed by the observation of quantum oscillations (Vignolle {\it et al.} 2008), persist.

The inferred correlation between $\alpha_1(0)$ and $T_c$ beyond $p_{\rm crit}$ agrees with an earlier ADMR study of $\Gamma$($T$,{\bf k}) in T2201 (Abdel-Jawad {\it et al.} 2007) and has also been seen more recently in the quasi-one-dimensional organic superconductor (TMTSF)$_2$X (X = PF$_6$, ClO$_4$) (Doiron-Leyraud {\it et al.} 2009). This correlation, coupled with the striking similarity between the angle-dependence of $\gamma_{\rm aniso}$ and the order parameter symmetry, implies that the interaction responsible for this anisotropic $T$-linear scattering rate is also involved in the superconducting pairing. Collectively, these numerous features (the additive components to $\Gamma$($T$, {\bf k}), the form of the anisotropy in $\gamma_{\rm aniso}$, the vanishing of the $T$-linear component along the nodes, its persistence to low temperatures and across the overdoped regime and its correlation with $T_c$) all place strong constraints on the development of any theoretical model put forward to explain the pairing mechanism and charge dynamics of hole-doped cuprates.

A $T$-linear scattering rate is often indicative of scattering off a bosonic mode. Obvious candidates in the cuprates include phonons, $d$-wave pairing fluctuations (Ioffe \& Millis 1998), spin (Stojkovic \& Pines 1997) and charge (Castellani {\it et al.} 1995) fluctuations. Since all, bar phonons, appear to vanish in heavily overdoped non-superconducting cuprates however (Wakimoto {\it et al.} 2004, Reznik {\it et al.} 2006), it is difficult to single one out at this stage. For a bosonic mode to be the source of $T$-linear scattering, the continuation of its linear $T$-dependence to very low temperatures is highly constraining, requiring as it does the presence of an extremely low energy scale. A robust $T$-linear scattering rate arises, for example, in bipolaron (Alexandrov 1997) or other two-dimensional (2D) boson-fermion mixtures (Wilson 2008, Chakraborty \& Phillips 2009) due to fermions scattering off density fluctuations of charge 2e bosons, though how this scattering persists beyond $p_{\rm crit}$ and the closing of the pseudogap is not yet clear.

An alternative origin for this {\bf k}-space anisotropic scattering is real-space (correlated) inhomogeneity. Indeed, STM measurements on hole-doped cuprates have found evidence for intense anisotropic scattering with a linear energy dependence (Alldredge {\it et al.} 2008) associated with an inhomogeneous superconducting state. The linear $T$-dependence of $\gamma_{\rm aniso}(T)$ and its angular dependence are also mirrored in the frequency-dependent {\it single}-particle scattering rate $\Gamma(\omega)$ inferred from ARPES studies on LSCO (Chang {\it et al.} 2008), suggesting that the two derive from the same origin. This combined linearity in both the temperature and frequency scales is typically referred to as marginal Fermi-liquid (MFL) phenomenology (Varma {\it et al.} 1989) though its preservation over such a wide doping range is inconsistent with models based on conventional quantum criticality. Indeed, problems with the application of such single parameter scaling hypotheses to the in-plane resistivity of high-$T_c$ cuprates is already well documented (Phillips \& Chamon 2005).

Assuming an effective interaction with the appropriate $d$-wave form factor, the transport scattering rate and electrical resistivity of a 2D metal close to a Pomeranchuk instability was recently shown to follow a $T^{4/3}$ dependence as $T \rightarrow 0$ (Dell'Anna \& Metzner 2007, 2009). Whilst this gives rise to a scattering rate (and resistivity) with a form approximately similar to that given in Eqn. (1) (when combined with impurity scattering and conventional, isotropic electron-electron scattering), there is no implicit correlation between the strength of the $d$-wave scattering and $T_c$ within this model.  Finally, in recent functional renormalization group calculations for a 2D Hubbard model, Ossadnik and co-workers have uncovered a strongly angle-dependent $T$-linear scattering term that originates from spin fluctuation vertex corrections (Ossadnik {\it et al.} 2008). Significantly, this $T$-linear scattering rate shows strong doping dependence too and vanishes as the superconductivity disappears on the overdoped side, in good agreement with experimental observations (Abdel-Jawad {\it et al.} 2007).

The high temperature $T$-linear coefficient $\alpha_1(\infty)$ shows a strikingly different doping dependence to $\alpha_1(0)$. Above $p = p_{\rm crit}$, $\alpha_1(\infty)$ attains a constant value of $\sim 1.0 \pm 0.1$ $\mu\Omega$cm/K. Such insensitivity to doping leads us to examine whether $\alpha_1(\infty)$ represents some sort of fundamental limit. For a 2D Drude metal,

\begin{equation}
{\rm d}\rho_{ab}/{\rm d}T = (2\pi \hbar d/e^2v_Fk_F) {\rm d}(1/\tau)/{\rm d}T
\end{equation}

\noindent
where $d$ is the interlayer spacing (= 0.64 nm in LSCO), $v_F$ is the Fermi velocity and $k_F$ the Fermi wave vector. Taking typical ($p$-independent) values of $v_F$ (= 8.0 x 10$^4$ ms$^{-1}$) and $k_F$ (= 7 nm$^{-1}$) for the anti-nodal states in overdoped LSCO (Yoshida {\it et al.} 2007) we find that for $\alpha_1(0)$ = $1 \mu\Omega$cm/K, the momentum-averaged scattering rate $\hbar/\tau  \sim \pi k_BT$. Given that the anisotropic scattering rate varies as cos$^22\varphi$ within the plane ($\varphi$ being the angle between the $k$-vector and the Cu-O-Cu bond direction) (Abdel-Jawad {\it et al.} 2006), this is equivalent to $\hbar/\tau  = 2\pi k_BT$ for states near ($\pi$, 0). Intriguingly, this intense level of scattering corresponds to the so-called \lq Planckian dissipation limit' beyond which Bloch-wave propagation becomes inhibited, i.e. the quasiparticle states themselves become incoherent (Zaanen 2004).

Related to this, ARPES experiments on overdoped cuprates have shown that the onset of $T$-linear resistivity coincides with the loss of in-plane quasiparticle coherence at the anti-nodal points near ($\pi$, 0), manifest in the disappearance of a peak in the ARPES spectral function (Kaminski {\it et al.} 2003). Taken together, these two features imply that the $T$-linear resistivity in overdoped cuprates is a signature of quasiparticle incoherence once the scattering rate near the zone boundary exceeds $2\pi k_BT$. Accordingly, we label the onset of $T$-linear resistivity beyond $p_{\rm crit}$ in Fig. 1 as $T_{\rm coh}$, the onset temperature of anti-nodal coherence (with decreasing temperature).

Below $p$ = 0.19, $\alpha_1(\infty)$ starts to rise sharply with decreasing Sr content. Correspondingly, the ratio $\alpha_1(0)$/$\alpha_1(\infty)$, plotted in Fig. 3b, maximizes at $p_{\rm crit}$ at a value close to unity. This maximum is a clear signature of a fundamental change in the quasiparticle response, coincident with the opening of the anisotropic $d$-wave pseudogap as determined by bulk thermodynamic measurements (Tallon \& Loram 2001). It is worth noting however that this doping level is also the point at which, according to ARPES (Yoshida {\it et al.} 2007), the Fermi level crosses a van Hove singularity in LSCO. Whilst this may have some impact on the transport properties (Gor'kov \& Teitelbaum 2006), band-structure calculations have shown that van Hove singularities in cuprates are very system dependent, while the resistivity behaviour is known to be quantitatively universal amongst the entire cuprate family (Hussey 2008).

Within the present picture, the rise in $\alpha_1(\infty)$ below $p_{\rm crit}$ is interpreted as a manifestation of the growth of the incoherent region (i.e. as defined here by the condition $\hbar/\tau \geq 2\pi k_BT$) away from ($\pi$, 0) as the effective interaction continues to intensify with decreasing $x$. As $T$ is lowered, those same incoherent regions become gapped out leading to a progressive destruction of the large Fermi surface found above $p_{\rm crit}$ either into Fermi arcs centred along the zone diagonals (Norman {\it et al.} 1998) or into electron and hole pockets (LeBoeuf {\it et al.} 2007). For the anisotropic coefficient $\alpha_1(0)$ however, the reduction in the total number of coherent states below $p_{\rm crit}$ is more than offset by the removal of the strong scattering sinks near the zone boundary, leading to an overall decrease, or at least a saturation, in $\alpha_1(0)$ with further reduction in carrier density.

\begin{figure}
\begin{center}
\includegraphics [width=11.0cm, bb=0 0 519 400]{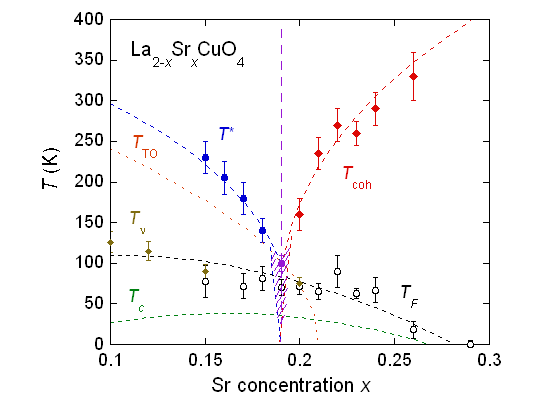}
\caption {Temperature vs. doping phase diagram of La$_{2-x}$Sr$_x$CuO$_4$ as extracted from the temperature derivative of $\rho_{ab}(T)$. As in Fig. 1, the labels $T_F$, $T_{\rm TO}$, $T\ast$ and $T_{\rm coh}$ represent respectively the onset temperature for short-lived vortex excitations, the tetragonal to orthorhombic structural transition, the opening of the pseudogap and the onset of quasi-particle incoherence. $T_{\nu}$ represents the onset of a positive Nernst signal in La$_{2-x}$Sr$_x$CuO$_4$ (Xu {\it et al.} 2000) at temperatures above the $T_c$ parabola. The dashed lines are all guides to the eye.} \label{Figure4}
\end{center}
\end{figure}

The doping dependence of the various temperature scales extracted from d$\rho_{ab}$/d$T(T)$ is captured in Figure 4. The labels $T_F$, $T_{\rm TO}$, $T^{\ast}$ and $T_{\rm coh}$ refer to respectively the marked upturn in d$\rho_{ab}$/d$T$ above $T_c$, the tetragonal to orthorhombic structural transition, the downturn in d$\rho_{ab}$/d$T$ for $p < 0.19$ and corresponding downturn above $p_{\rm crit}$. $T_{\nu}$ represents the onset of a positive Nernst signal seen in La$_{2-x}$Sr$_x$CuO$_4$ (Xu {\it et al.} 2000). As mentioned above, at doping levels where such measurements have been performed, $T_F$ coincides with $T_{\nu}$, affirming that the sharp downturn in $\rho_{ab}(T)$ (upturn in d$\rho_{ab}$/d$T$) marks the onset of short-lived vortex excitations existing well above the bulk superconducting transition. What is new in this work is the observation that this anomalous phase fluctuation regime persists well into the overdoped regime. In LSCO24, for example, $T_F \sim 3T_c$. Given that the original picture of phase fluctuating superconductivity in cuprates was applied to the heavily underdoped region where the superfluid density is small (Emery \& Kivelson 1995), this result is rather surprising and clearly warrants further attention.

The effect of the tetragonal to orthorhombic structural transition on d$\rho_{ab}$/d$T$ at $T$ = $T_{\rm TO}$ is small but well-defined (see Fig. 1b). It is manifest similarly in the double derivative analysis of $\rho_{ab}(T)$ performed by Ando and co-workers (Ando {\it et al.} 2004) though comparison with actual structural analysis data (Yamada {\it et al.} 1998) informs that the dashed line depicting $T^{\ast}$ in Fig. 2(c) of their paper is in fact $T_{\rm TO}$.

According to the literature, the standard definition of $T^{\ast}$ is the temperature below which $\rho_{ab}(T)$ starts to deviate from its linear-$T$ behaviour at high temperature (Daou {\it et al.} 2009). Different methods for extracting $T^{\ast}$ often lead to markedly different values, though derivative plots are invariably the most sensitive and therefore return the highest $T^{\ast}$ values (Hussey {\it et al.} 1997). The above definition can be deceiving however, particularly in LSCO where the deviation in d$\rho_{ab}$/d$T$ is upward. As illustrated in Fig. 1, the form of $\rho_{ab}(T)$ is remarkably similar on both sides of $p_{\rm crit}$ making it difficult to distinguish between pseudogap opening and the coherent/incoherent crossover in that region of the phase diagram. Indeed, the only way these two temperature scales can be distinguished is via their doping dependencies: whilst $T^{\ast}$ decreases with increasing $p$, $T_{\rm coh}$ shows the opposite trend. Within our experimental uncertainty, it is not yet possible to determine whether $T^{\ast}$ and $T_{\rm coh}$ vanish or simply cross around $p$ = $p_{\rm crit}$ - a detailed study of more closely-spaced doping levels will be required to address this point. Nevertheless, the fact that the ratio  $\alpha_1(0)$/$\alpha_1(\infty)$ depicted in Fig. 3b maximizes at a value close to one suggests strongly that $p$ = 0.19 is the point at which both temperature scales vanish. The shaded area in Fig. 4 serves to reflect this uncertainty.

The location of $T_{\rm coh}(p)$ for LSCO from our analysis is found to be in excellent agreement with that found for overdoped Bi2212 (Kaminski {\it et al.} 2003), overdoped Tl2201 and Ca-doped YBa$_2$Cu$_3$O$_{7-\delta}$ (Naqib {\it et al.} 2003), affirming that it is a generic feature of all overdoped cuprates. Above this line, the anti-nodal quasiparticles are incoherent (Kaminski {\it et al.} 2003). In a conventional metal, $\rho(T)$ tends to saturate at high temperatures once the inelastic scattering rate exceeds the effective bandwidth - the so-called Mott-Ioffe-Regel (MIR) limit. In cuprates however, as in other \lq bad' metals (Hussey {\it et al.} 2004), the onset of incoherence is accompanied by a loss of low-frequency spectral weight to frequencies of order the on-site Coulomb repulsion $U$ (Merino {\it et al.} 2000, Takenaka {\it et al.} 2003). This loss of low-frequency spectral weight manifests itself as a dip in the dc conductivity and thus a rise in the electrical resistivity to values beyond the MIR limit. A connection between this spectral weight reduction and the preservation of $T$-linear resistivity has yet to be firmly established, though a phenomenological model based on a dominant, anisotropic, $T^2$ scattering rate and proximity to the MIR limit demonstrated a link between the onset of $T$-linearity in cuprates and the loss of coherence at the anti-nodes (Hussey 2003). Other, more recent proposals attribute the high-temperature $T$-linear resistivity to incoherent transport induced either by scattering off gauge fluctuations (Senthil \& Lee 2009) or by the motion of hard-core charged bosons (Lindner \& Auerbach 2009).

Finally, before concluding, let us consider briefly the quadratic term in $\rho_{ab}(T)$. According to ARPES,  the single particle scattering rate $\Gamma(\omega)$ of overdoped and optimally doped cuprates is also (approximately) quadratic along the nodal directions (Kordyuk {\it et al.} 2004, Koralek {\it et al.} 2006). This combined quadratic temperature and frequency dependence confirms electron-electron (Umklapp) scattering as its origin. As can be seen in the derivatives plots of Fig. 1 however, the $T^2$ term extends in some cases up to $T^\ast$ or $T_{\rm coh}$. Whilst this may seem surprising, given that it is a significant fraction of the Fermi energy, it is not uncommon, particularly in correlated transition metal oxides (Hussey 2005).

\section{Conclusions}

A $T$-linear resistivity, persisting in some cases up to 1000 K (Gurvitch \& Fiory 1987, Ono {\it et al.} 2007), has been one of the defining characteristics of the normal state of hole-doped cuprates, yet despite sustained theoretical efforts over the past two decades, its origin and its relation to the superconducting mechanism remain a profound, unsolved mystery. In this article, we have analyzed the temperature derivative of in-plane resistivity data on a series of closely-spaced LSCO single crystals and have uncovered the presence of not one, but two seemingly independent $T$-linear coefficients that persist for all dopings 0.15 $\leq x \leq$ 0.3. Combining the results summarized in Figures 3 and 4, a coherent picture begins to emerge in which the two coefficients reflect different aspects of the normal state, namely the low-energy effective interaction and quasiparticle decoherence.

For $x > 0.3$, LSCO is a correlated Fermi-liquid with a large, $T^2$ resistivity extending over a broad temperature range (Nakamae {\it et al.} 2003, 2009). As the carrier number falls and the system begins to track back towards the Mott insulating state at $p$ = 0, an additional anisotropic interaction develops, giving rise to the $T$-linear scattering term. The same interaction also drives up $T_c$, the condensation energy and superfluid density, the latter reaching a maximum at $p = p_{\rm crit}$ (Panagopoulos {\it et al.} 2003). At this point however, scattering intensity becomes so strong that those states near ($\pi$, 0) begin to de-cohere and given the proportionality with temperature, {\it decoherence occurs at all finite temperatures}, not just below $T^{\ast}$. With further reduction in doping, the strength of the interaction continues to rise, causing yet more states to lose coherence and leading to an overall reduction or saturation in $\alpha_1(0)$, a rise in $\alpha_1(\infty)$  and a corresponding suppression in the condensation energy and superfluid density.

Within this picture, the strength of the condensate is destroyed below $p_{\rm crit}$ because the interaction that promotes superconductivity ultimately destroys the very quasiparticles needed to form the condensate. It is tempting to speculate that the pseudogap also forms in response to this intense scattering, the electronic ground state lowering its energy through gapping out the incoherent, highly energetic anti-nodal states and in so doing, preventing scattering of the remnant quasiparticle states into those same regions. This implies then that, contrary to much current thinking, the physics behind pseudogap formation is already prevalent in the strongly overdoped regime, with decoherence, rather than competing order, being the principal driver. Whilst the two of course may be intimately linked, the absence of conventional, critical scaling behaviour in both the thermodynamic and transport properties of cuprates is at odds with the notion that $p_{\rm crit}$ defines some sort of zero-temperature phase transition.

Finally, concerning the phase diagram of hole-doped cuprates, we have shown evidence from $\rho_{ab}(T)$ data that $T^{\ast}$ and $T_F$ cross close to $p$ = 0.19. This coincidence probably explains why certain probes, particularly those that utilize $c$-axis tunneling, invariably support scenarios in which $T^{\ast}$ tracks the superconducting dome on the overdoped side since they cannot distinguish between the loss of states due to pseudogap formation and the loss of states due to superconducting phase fluctuations. Perhaps future reviews of pseudogap phenomena will recognize this distinction and allow the debate to move on to the important issue of its fundamental origin.

\ack We would like to thank O. J. Lipscombe, S. M. Hayden, Y. Tanabe, T. Adachi, Y. Koike, M. Nohara, H. Takagi and A. P. Mackenzie for the single crystal samples. This work was supported by the EPSRC (UK), the Royal Society, LNCMI-T, the French ANR IceNET and EuroMagNET.

\section*{References}

\noindent
Abdel-Jawad, M., Kennett, M. P., Balicas, L., Carrington, A., Mackenzie, A. P., McKenzie, R. H. \& Hussey, N. E. 2006 Anisotropic scattering and anomalous normal-state transport in a high-temperature superconductor {\it Nature Phys.} {\bf 2}, 821-824. (DOI:10.1038/nphys449 Letter)

Abdel-Jawad, M., Balicas, L., Carrington, A., Charmant, J. P. H., French, M. M. J. \& Hussey, N. E. 2007 Correlation between the superconducting transition temperature and anisotropic quasiparticle scattering in Tl$_2$Ba$_2$CuO$_{6+\delta}$ {\it Phys. Rev. Lett.} {\bf 99}, 107002. (DOI:10.1103/PhysRevLett.99.107002)

Alexandrov, A. S. 1997 Condensation of charged bosons hybridised with fermions {\it Physica} {\bf 274C}, 237-247. (DOI:10.1016/S0921-4534(96)00673-9)

Alldredge, J. W., Lee, J., McElroy, K., Wang, M., Fujita, K., Kohsaka, Y., Taylor, C., Eisaki, H., Uchida, S., Hirschfeld, P. J. \& Davis, J. C. 2008 Evolution of the electronic excitation spectrum with strongly diminishing hole density in superconducting Bi$_2$Sr$_2$CaCu$_2$O$_{8+delta}$ {\it Nature Phys.} {\bf 4}, 319-326. (DOI:10.1038/nphys917 Article)

Ando, Y., Komiya, S., Segawa, K., Ono, S. \& Kurita, Y. 2004 Electronic phase diagram of high-$T_c$ cuprate superconductors
from a mapping of the in-plane resistivity curvature {\it Phys. Rev. Lett.} {\bf 93}, 267001. (DOI:10.1103/PhysRevLett.93.267001)

Castellani, C., di Castro, C. \& Grilli, M. 1995 Singular quasiparticle scattering in the proximity of charge instabilities, {\it Phys. Rev. Lett.} {\bf 75}, 4650-4653. (DOI:10.1103/PhysRevLett.75.4650)

Chakraborty, S. \& Phillips, P. 2009 Two-fluid model of the pseudogap of high-temperature cuprate superconductors based on charge-2e bosons {\it Phys. Rev. B} {\bf 80}, 132505. (DOI:10.1103/PhysRevB.80.132505)

Chang, J., Shi, M., Pailh\'{e}s, S., M\"{a}nsson, M., Claesson, T., Tjernberg, O., Bendounan, A., Sassa, Y., Patthey, L., Momono, N., Oda, M., Ido, M., Guerrero, S., Mudry, C. \& Mesot, J. 2008 Anisotropic quasiparticle scattering rates in slightly underdoped to optimally doped high-temperature La$_{2-x}$$_x$CuO$_4$ superconductors{\it Phys. Rev. B} {\bf 78}, 205103. (DOI:10.1103/PhysRevB.78.205103)

Cooper, R. A., Wang, Y., Vignolle, B., Lipscombe, O. J., Hayden, S. M., Tanabe, Y., Adachi, T., Koike, Y., Nohara, M., Takagi, H., Proust, C. \&
Hussey, N. E. 2009 Anomalous criticality in the electrical resistivity of La$_{2–x}$Sr$_x$CuO$_4$ {\it Science} {\bf 323}, 603-607. (DOI:10.1126/science.1165015)

Custers, J., Gegenwart, P., Wilhelm, H., Neumaier, K., Tokiwa, Y., Trovarelli, O., Geibel, C., Steglich, F., P\'{e}pin, C. \& Coleman, P. 2003 The break-up of heavy electrons at a quantum critical point {\it Nature} {\bf 424}, 524-527. (DOI:10.1038/nature01774)

Daou, R., Doiron-Leyraud, N., LeBoeuf, D., Li, S. Y., Lalibert\'{e}, F., Cyr-Choini\`{e}re, O., Jo, Y. J., Balicas, L., Yan, J.-Q., Zhou, J.-S., Goodenough, J. B. \&  Taillefer, L. 2009 Linear temperature dependence of resistivity and change in the Fermi surface at the pseudogap critical point of a high-$T_c$ superconductor {\it Nature Phys.} {\bf 5}, 31-34. (DOI:10.1038/nphys1109 Letter)

Dell'Anna, L. \& Metzner, W. 2007 Electrical resistivity near Pomeranchuk instability in two dimensions {\it Phys. Rev. Lett.} {\bf 98}, 136402. (DOI:10.1103/PhysRevLett.98.136402)

Dell'Anna, L. \& Metzner, W. 2009 Erratum: Electrical resistivity near Pomeranchuk instability in two dimensions [Phys. Rev. Lett. 98, 136402 (2007)] {\it Phys. Rev. Lett.} {\bf 103}, 159904. (DOI: 10.1103/PhysRevLett.103.159904)

Doiron-Leyraud, N., Auban-Senzier, P., Ren\'{e} de Cotret, S., Sedeki, A., Bourbonnais, C., J\'{e}rome, D., Bechgaard, K. \& Taillefer, L. 2009 Correlation between linear resistivity and $T_c$ in organic and pnictide superconductors (arXiV:condmat0905.0964v1).

Emery, V. \& Kivelson, S. 1995 Importance of phase fluctuations in superconductors with small superfluid density {\it Nature} {\bf 374}, 434–437. (DOI:10.1038/374434a0)

Fauqu\'{e}, B., Sidis, Y., Hinkov, V., Pailh\'{e}s, S., Lin, C. T., Chaud, X. \& Bourges, P. 2006 {\it Phys. Rev. Lett.} {\bf 96}, 197001. (DOI:10.1103/PhysRevLett.96.197001)

French, M. M. J. \& Hussey, N. E. 2009 Orbital origin of field-induced \lq quantum criticality' in overdoped Tl$_2$Ba$_2$CuO$_{6+\delta}$  {\it. Proc. Nat. Acad. Sci.} {\bf 105}, E58. (DOI:10.1073Bpnas.0805887105)

French, M. M. J., Analytis, J. G., Carrington, A., Balicas, L. \& Hussey, N. E. 2009 Tracking anisotropic scattering in overdoped Tl$_2$Ba$_2$CuO$_{6+\delta}$ above 100K {\it New J. Phys.} {\bf 11}, 055057. (DOI:10.1088/1367-2630/11/5/055057)

Gor'kov, L. P. \& Teitelbaum, G. B. 2006 Interplay of externally doped and thermally activated holes in La$_{2-x}$$_x$CuO$_4$ and their impact on the pseudogap crossover {\it Phys. Rev. Lett.} {\bf 97}, 247003. (DOI:10.1103/PhysRevLett.97.247003)

Gurvitch, M. \& Fiory, A. T. 1987 Resistivity of La$_{2-x}$$_x$CuO$_4$ and YBa$_2$Cu$_3$O$_7$ to 1100K: Absence of saturation and its implications {\it Phys. Rev. Lett.} {\bf 59}, 1337-1340. (DOI:10.1103/PhysRevLett.59.1337)

H\"{u}fner, S., Hossain, M. A., Damascelli, A. \& Sawatsky, G. A. 2008 Two gaps make a high-temperature superconductor? {\it Rep. Prog. Phys.} {\bf 71}, 062501. (DOI:10.1088/0034-4885/71/6/062501)

Hussey, N. E., Nozawa, K., Takagi, H., Adachi, S. \& Tanabe, K. 1997 Anisotropic resistivity of YBa$_2$Cu$_4$O$_8$: Incoherent-to-metallic crossover
in the out-of-plane transport {\it. Phys. Rev. B} {\bf 56}, R11423-R11426. (DOI:10.1103/PhysRevB.56.R11423)

Hussey, N. E. 2003 The normal state scattering rate in high-$T_c$ cuprates {\it Eur. Phys. J. B} {\bf 31}, 495-507. (DOI:10.1140/epjb/e2003-00059-9)

Hussey, N. E., Takenaka, K. \& Takagi, H. 2004 Universality of the Mott–Ioffe–Regel limit in metals {\it Phil. Mag.} {\bf 84}, 2847-2864. (DOI:10.1080/14786430410001716944)

Hussey, N. E. 2005 Non-generality of the Kadowaki-Woods ratio in correlated oxides {\it J. Phys. Soc. Japan} {\bf 104}, 1107-1110. (DOI:10.1143/JPSJ.74.1107)

Hussey, N. E. 2008 Phenomenology of the normal state in-plane transport properties of high-$T_c$ cuprates {\it J. Phys. - Condens. Matt.} {\bf 20}, 123201. (DOI:10.1088/0953-8984/20/12/123201)

Ioffe, L. B. \& Millis, A. J. 1998 Zone-diagonal-dominated transport in high-$T_c$ cuprates {\it Phys. Rev. B} {\bf 58}, 11631-11637. (DOI:10.1103/PhysRevB.58.11631)

Kaminski, A., Rosenkranz, A., Fretwell, H. M, Li, Z. Z., Raffy, H., Randeria, M., Norman, M. R. \& Campuzano, J. C. 2003 Crossover from coherent to incoherent electronic excitations in the normal state of Bi$_2$Sr$_2$CaCu$_2$O$_{8+\delta}$ {\it Phys. Rev. Lett.} {\bf 90}, 207003. (DOI:10.1103/PhysRevLett.59.1337)

Kopp, A., Ghosal, A. \& Chakravarty, S. 2007 Competing ferromagnetism in high-temperature copper oxide superconductors {\it Proc. Natl. Acad. Sci.} {\bf 104}, 6123. (DOI:10.1073/pnas.0701265104)

Koralek, J. D., Douglas, J. F., Plumb, N. C., Sun, Z., Fedorov, A. V., Murnane, M. M., Kapteyn, H. C., Cundiff, S. T., Aiura, Y., Oka, K., Eisaki, H. \& Dessau, D. S. 2006 Laser based angle-resolved photoemission, the sudden approximation, and quasiparticle-like spectral peaks in Bi$_2$Sr$_2$CaCu$_2$O$_{8+\delta}$ {\it Phys. Rev. Lett.} {\bf 96} 017005. (DOI:10.1103/PhysRevLett.96.017005)

Kordyuk, A. A., Borisenko, S. V., Koitzsch, A., Fink, J., Knupfer, M., Buchner, B., Berger, H., Margaritondo, G., Lin, C. T., Keimer, B., Ono, S. \& Ando, Y. 2004 Manifestation of the magnetic resonance mode in the nodal quasiparticle lifetime of the superconducting cuprates {\it Phys. Rev. Lett.} {\bf 92} 257006. (DOI:10.1103/PhysRevLett.92.257006)

LeBoeuf, D., Doiron-Leyraud, N., Levallois, J., Daou, R., Bonnemaison, J.-B., Hussey, N. E., Balicas, L., Ramshaw, B. J., Liang, R., Bonn, D. A., Hardy, W. N., Adachi, S., Proust, C. \& Taillefer, L. 2007 Electron pockets in the Fermi surface of hole-doped high-$T_c$ superconductors {\it Nature} {\bf 450}, 533-536. (DOI:10.1038/nature06332)

Lindner, N. H. \& Auerbach, A. \lq Bad metal' conductivity of hard core bosons (arXiV:0910.4158v1).

Mackenzie, A. P., Julian, S. R., Sinclair, D. C. \& Lin, C. T. 1996 Normal-state magnetotransport in superconducting Tl$_2$Ba$_2$CuO$_{6+\delta}$ to millikelvin temperatures {\it Phys. Rev. B} {\bf 53}, 5848-5855. (DOI: 10.1103/PhysRevB.53.5848)

Manako, T., Kubo, Y. \& Shimakawa, Y. 1992 {\it Phys. Rev. B} {\bf 46}, 11019-11024. (DOI:10.1103/PhysRevB.46.11019)

Merino, J. \& McKenzie, R. H. 2000 Transport properties of strongly correlated metals: A dynamical mean-field approach {\it Phys. Rev B} {\bf 61}, 7996-8008. (DOI:10.1103/PhysRevB.61.7996)

Naqib, S. H., Cooper, J. R., Tallon, J. L. \& Panagopoulos, C. 2003 Temperature dependence of electrical resistivity of high-$T_c$
cuprates – from pseudogap to overdoped regions {\it Physica} {\bf 387C}, 365-372. (DOI:10.1016/S0921-4534(02)02330-4)

Nakamae, S., Behnia, K., Yates, S. J. C., Mangkorntong, N., Nohara, M., Takagi, H. \& Hussey, N. E. 2003 Electronic ground state of heavily overdoped nonsuperconducting La$_{2-x}$Sr$_x$CuO$_4$  {\it Phys. Rev. B} {\bf 68}, 100502(R). (DOI: 10.1103/PhysRevB.68.100502)

Nakamae, S., Behnia, K., Mangkorntong, N., Nohara, M., Takagi, H., Yates, S. J. C. \& Hussey, N. E. 2009 Erratum: Electronic ground state of heavily overdoped nonsuperconducting La$_{2-x}$Sr$_x$CuO$_4$ [Phys. Rev. B 68, 100502 (2003)] {\it Phys. Rev. B} {\bf 79}, 219904(E).  (DOI:10.1103/PhysRevB.79.219904)

Norman, M. R., Ding, H., Randeria, M., Campuzano, J. C., Yokoya, T., Takeuchi, T., Takahashi, T., Mochiku, T., Kadowaki, K., Guptasarma P. \& Hink, D. G. 1998 Destruction of the Fermi surface in underdoped high-$T_c$ superconductors {\it Nature} {\bf 392}, 157–160. (DOI:10.1038/32366)

Norman, M. R., Pines, D. \& Kallin, C. 2005 The pseudogap: Friend or foe of high-$T_c$? {\it Adv. Phys.} {\bf 54}, 715-733. (DOI:10.1080/00018730500459906)

Ono, S., Komiya, S. \& Ando, Y. 2007 Strong charge fluctuations manifested in the high-temperature Hall coefficient of high-$T_c$ cuprates {\it Phys. Rev. B} {\bf 75}, 024515. (DOI:10.1103/PhysRevB.75.024515)

Ossadnik, M., Honerkamp, C., Rice, T. M. \& Sigrist, M. 2008 Breakdown of Landau theory in overdoped cuprates near the onset of superconductivity {\it Phys. Rev. Lett.} {\bf 101}, 256405. (DOI:10.1103/PhysRevLett.101.256405)

Paglione, J., Tanatar, M. A., Hawthorn, D. G., Boaknin, E., Hill, R. W., Ronning, F., Sutherland, M., Taillefer, L., Petrovic, C. \& Canfield, P. C. 2003 Field-induced quantum critical point in CeCoIn$_5$ {\it Phys. Rev. Lett.} {\bf 91}, 246405. (DOI:10.1103/PhysRevLett.91.246405)

Panagopoulos, C., Xiang, T., Anukool, W., Cooper, J. R., Wang, Y. S. \& Chu, C. W. 2003 Superfluid response in monolayer high-$T_c$ cuprates {\it Phys. Rev. B} {bf 67}, 220502. (DOI:10.1103/PhysRevB.67.220502)

Phillips, P. \& Chamon, C. 2005 Breakdown of one-parameter scaling in quantum critical scenarios for high-temperature copper-oxide superconductors {\it Phys. Rev. Lett.} {\bf 95}, 107002. (DOI:10.1103/PhysRevLett.95.107002)

Proust, C., Boaknin, E., Hill, R. W., Taillefer, L. \& Mackenzie, A. P. 2002 Heat transport in a strongly overdoped cuprate: Fermi liquid
and a pure $d$-wave BCS superconductor {\it Phys. Rev. Lett.} {\bf 89}, 147003. (DOI:10.1103/PhysRevLett.89.147003)

Reznik, D., Pintschovius, L., Ito, M., Iikubo, S., Sato, M., Goka, H., Fujita, M., Yamada, K., Gu, G. D. \& Tranquada, J. M. 2006 Electron-phonon coupling reflecting dynamic charge inhomogeneity in copper-oxide superconductors {\it Nature} {\bf 440}, 1170-1173. (DOI:10.1038/nature04704)

Senthil, T. \& Lee, P. A. 2009 Coherence and pairing in a doped Mott insulator: application to the cuprates  {\it Phys. Rev. Lett.} {\bf 103}, 076402. (DOI:10.1103/PhysRevLett.103.076402)

Shibauchi, T., Krusin-Elbaum, L., Hasegawa, M., Kasahara, Y., Okazaki, R. \& Matsuda, Y. 2008 Field-induced quantum critical route to a Fermi liquid
in high-temperature superconductors {\it Proc. Nat. Acad. Sci.} {\bf 105}, 7120-7123. \overleftarrow{}(DOI:10.1073Cpnas.0712292105)

Sonier, J. E., Kaiser, C. V., Pacradouni, V., Sabok-Sayr, S. A., Cochrane, C., MacLaughlin, D. E., Komiya, S. \& Hussey, N. E. 2009 Emergence of a novel frozen magnetic state in a heavily overdoped non-superconducting copper oxide (arXiV:condmat0911.0407v1).

Stojkovic, B. P. \& Pines, D. 1997 Theory of the longitudinal and Hall conductivities of the cuprate superconductors {\it Phys. Rev. B} {\bf 55}, 8576-8595. (DOI:10.1103/PhysRevB.55.8576)

Takenaka, K., Nohara, J., Shiozaki, R. \& Sugai, S. 2003 Incoherent charge dynamics of La$_{2-x}$Sr$_x$CuO$_4$: Dynamical localization and resistivity saturation {\it Phys. Rev. B} {\bf 68}, 134501. (DOI:10.1103/PhysRevB.68.134501)

Tallon, J. L. \& Loram, J. W. 2001 The doping dependence of T$^\ast$ - what is the real high-$T_c$ phase
diagram? {\it Physica} {\bf 349C}, 53-68. (DOI:10.1016/S0921-4534(00)01524-0)

Varma, C. M., Littlewood, P. B., Schmitt-Rink, S., Abrahams, E., \& Ruckenstein, A. E. 1989 Phenomenology of the normal state of Cu-O high-temperature superconductors {\it Phys. Rev. Lett.} {\bf 63}, 1996-1999. (DOI:10.1103/PhysRevLett.63.1996)

Vignolle, B., Carrington, A., Cooper, R. A., French, M. M. J., Mackenzie, A. P., Jaudet, C., Vignolles, D., Proust, C. \& Hussey, N. E. 2008 Quantum oscillations in an overdoped high-$T_c$ superconductor {\it Nature} {\bf 455}, 952-955. (DOI:10.1038/nature07323)

Wakimoto, S., Zhang, H., Yamada, K., Swainson, I., Kim, H. \& Birgeneau, R. J. 2004 Direct relation between the low-energy spin excitations and superconductivity of overdoped high-$T_c$ superconductors {\it Phys. Rev. Lett.} {\bf 92}, 217004. (DOI:10.1103/PhysRevLett.92.217004)

Wilson, J. A. 2008 Elucidation of the origins of HTSC transport behaviour and quantum oscillations (arXiV:condmat0811.3096v2).

Xia, J., Schemm, E., Deutscher, G., Kivelson, S. A., Bonn, D. A., Hardy, W. N., Liang, R., Siemons, W., Koster, G., Fejer, M. M. \& Kapitulnik, A. 2008 {\it Phys. Rev. Lett.} {\bf 100}, 127002. (DOI:10.1103/PhysRevLett.100.127002)

Xu, Z. A., Ong, N. P., Wang, Y., Kakeshita, T. \& Uchida, S. 2000 Vortex-like excitations and the onset of superconducting phase fluctuations in underdoped La$_{2-x}$Sr$_x$CuO$_4$ {\it Nature} {\bf 406}, 486-489. (DOI:10.1038/35020016)

Yamada, K., Lee, C. H., Kurahashi, K., Wada, J., Wakimoto, S., Ueki, S., Kimura, H., Endoh, Y., Hosoya, S., Shirane, G., Birgeneau, R. J., Greven, M., Kastner, M. A. \& Kim, Y. J. 1998 Doping dependence of the spatially modulated dynamical spin correlations and the superconducting-transition temperature in La$_{2-x}$Sr$_x$CuO$_4$ {\it Phys. Rev. B} {\bf 57}, 6165-6172. (DOI:10.1103/PhysRevB.57.6165)

Yoshida, T., Zhou, X. J., Lu, D. H., Komiya, S., Ando, Y., Eisaki, H., Kakeshita, T., Uchida, S., Hussain, Z., Shen, Z-X. \& Fujimori, A. 2007 Low-energy electronic structure of the high-$T_c$ cuprates La$_{2-x}$Sr$_x$CuO$_4$ studied by angle-resolved photoemission spectroscopy {\it J. Phys.: Condens. Matt.} {\bf 19}, 125209. (DOI:10.1088/0953-8984/19/12/125209)

Zaanen, J. 2004 Why the temperature is high {\it Nature} {\bf 430}, 512-513. (DOI:10.1038/430512a)

\end{document}